\documentclass[twocolumn,twoside,preprintnumbers,amsmath,amssymb,showkeys,showpacs,floatfix]{revtex4}

\usepackage{epsfig,graphicx,fancyhdr}

\newcommand{\bl}[1]{\begin{equation}\label{#1}}
\newcommand{\ee}{\end{equation}}
\newcommand{\be}{\begin{equation}}
\newcommand{\bea}{\begin{eqnarray}}
\newcommand{\eea}{\end{eqnarray}}

\newcommand{\rec}[1]{\frac{1}{#1}}
\newcommand{\td}[2]{\frac{\mathrm{d}{#1}}{\mathrm{d}{#2}}}
\newcommand{\z}[1]{\left({#1}\right)}
\newcommand{\sz}[1]{\left[{#1}\right]}
\renewcommand{\v}[1]{\mathbf{#1}}

\newcommand{\ch}{\mathrm{cosh}}
\newcommand{\sh}{\mathrm{sinh}}

\renewcommand{\r}[1]{(\ref{#1})}

\newcommand{\elte}{ELTE, E{\"o}tv{\"o}s Lor{\'a}nd University, H - 1117 Budapest, P{\'a}zm{\'a}ny P. s. 1/A, Hungary}
\newcommand{\kfki}{MTA KFKI RMKI, H-1525 Budapest 114, POBox 49, Hungary}
\newcommand{\ift}{Instituto de F\'\i sica Te\'orica - UNESP, Rua Pamplona 145, 01405-900 S\~ao Paulo, SP, Brazil}
\begin{document}
\title{Accelerating Solutions of Perfect Fluid Hydrodynamics for \\
Initial Energy Density and Life-Time Measurements in Heavy Ion Collisions
}

\author{T.~Cs{\"o}rg\H{o}}      \affiliation{\kfki}\affiliation{\ift}
\author{M.~I.~Nagy}        \affiliation{\elte}
\author{M.~Csan\'ad}    \affiliation{\elte}

\begin{abstract}
A new class of accelerating, exact, explicit and simple solutions of 
relativistic hydrodynamics is presented. Since these
new solutions yield a finite rapidity distribution, 
they lead to an advanced estimate of  the initial energy density and
life-time of high energy heavy ion collisions.
Accelerating solutions are also given for spherical expansions
in arbitrary number of spatial dimensions.
\keywords{relativistic hydrodynamics, exact solutions, initial energy density, particle spectra and correlations}
\end{abstract}

\pacs{25.75.-q,24.10.Nz,47.15.Hg}

\maketitle

\date{\today}
\thispagestyle{fancy}

\setcounter{page}{1}

\section{Introduction}
Relativistic hydrodynamics can be successfully applied both 
at the largest and the smallest scales of physics.
However, its equations are highly nonlinear, 
and only few exact relativistic solutions 
are presently known. 
We present here  a recently 
found family of exact solutions~\cite{Csorgo:2006ax}.
These generalize the Landau-Khalatnikov
solution~\cite{Landau:1953gs,Khalatnikov,Belenkij:1956cd},
as they are also accelerating and yield a finite rapidity
distribution. They are simple and explicit, similarly to 
the Hwa-Bjorken solution~\cite{Hwa:1974gn,Bjorken:1982qr}.
We show new 1+ 3 dimensional solutions~\cite{Csorgo:2006ax}, 
that are reduced to the exact hydro solutions of 
refs.~\cite{Csorgo:2003rt,Csorgo:2003ry,Biro:1999eh,Biro:2000nj,Sinyukov:2004am}
in the accelerationless limit.

\section{New simple solutions of relativistic hydrodynamics}
Let us denote the metric tensor by 
$g^{\mu\nu} = \mbox{\it diag}(1,-1,-1,-1)$, the four-velocity by
$u^{\mu}=\gamma (1, v \v{n})$, the pressure by  $p$, 
the energy density by $\varepsilon$, the
temperature by  $T$, the charged particle density by $n$
%\, 
%the chemical potential by $\mu$ 
and the entropy
density by $\sigma$. All fields depend on the coordinates $x^\mu$. 
We denote by $r$ the spatial coordinate $r_z$ in 1+1 dimensions, 
and the radial coordinate in 1+ $d$ dimensions (where $r\ge 0$).
Relativistic hydrodynamics 
%are the 
%relativistic Euler-equation, the energy conservation equation
%and the charge conservation:
expresses local momentum, energy and  charge conservation:
\bea
(\varepsilon+p)u^{\nu}\partial_{\nu}u^{\mu} &=& \z{g^{\mu\rho}-u^{\mu}u^{\rho}}\partial_{\rho}p , \label{Reul} \\
(\varepsilon + p)\partial_{\nu}u^{\nu}+u^{\nu}\partial_{\nu}\varepsilon & = & 0 , \label{RE}  \\
\partial_{\nu} (nu^{\nu}) & = & 0 . \label{e:cont}
\eea
From these relations entropy conservation
also follows: $\partial_{\nu} (\sigma u^{\nu})  =  0$.
Our choice of Equations of State (EoS) is that of an 
ideal gas: $p=nT$\,, $\varepsilon=\kappa p$.
For an ultra-relativistic ideal gas in $d$ spatial dimensions, 
$\kappa = d = 1/c_s^2$. 
In the considered case, $\sigma \propto T^{d}$.

We use the $\tau$ and $\eta$ Rindler coordinates:
\bea
 |r| < |t| : & r= \tau\sh\eta , & t= \pm \tau\ch\eta,  \\
 |r| > |t| : & r= \pm \tau\ch\eta , & t=\tau\sh\eta .
\eea
In order to generalize the Hwa-Bjorken solution, 
we introduced the $\lambda>0$ parameter,
and found the following class of solutions:
\bea
v  & = & \tanh\,\lambda\eta , \label{1dsolv}\\
p  & = & p_0\z{\frac{\tau_0}{\tau}}^{\lambda d (1+ 1/\kappa)}
	\left[\cosh\frac{\eta}{2}\right]^{
		-(d-1) \phi(\lambda) (1 + 1/\kappa) 
		}
	 ,  \label{1dsolp}\\
n  & = & n_0\z{\frac{\tau_0}{\tau}}^{\lambda d} \,
	\nu(s) \, 
	\left[\cosh\frac{\eta}{2}\right]^{-(d-1) \phi(\lambda)  }
	,  \label{1dsoln}\\
 T & = & T_0\z{\frac{\tau_0}{\tau}}^{\lambda d/\kappa}
	\, \rec{\nu(s)}\,
	\left[\cosh(\frac{\eta}{2})\right]^{
		-(d-1) \phi(\lambda) /\kappa 
		}
	. \label{1dsolT}
\eea
The initial pressure, number density and temperature are denoted by 
$p_0$, $n_0$ and $T_0$, respectively.
The above forms are hydrodynamical solutions for the 
special values of $\lambda $ and $\kappa$
as given below. 
The $\lambda=1$\,, $|r|<|t|$ , $\nu(s)=1$ case is the
 $d$-dimensional generalization of the well
known inertial Hwa-Bjorken flow~\cite{Hwa:1974gn,Bjorken:1982qr},
where $\kappa$ and $d > 1 $ are arbitrary. Here $\nu(s)$
is an arbitrary positive scaling function, given by the initial conditions,
and $s$ is  a scaling variable, that has a vanishing co-moving
derivative:
\bea
|r|<|t| & : & s(\tau,\eta)=\left(\frac{\tau_0}{\tau}\right)^{\lambda-1}
        \sinh\sz{\z{\lambda-1}\eta} , \label{1dsolsint}\\
|r|>|t| & : & s(\tau,\eta)=\left(\frac{\tau_0}{\tau}\right)^{\lambda-1}
    \cosh\sz{\z{\lambda-1}\eta}\label{1dsolsext} ,
\eea
valid for $\lambda\neq 1$. If $\lambda=1$, we have $s=s_0\eta$ for $|r|<|t|$, while $s=\tau/\tau_0$ for
$|r|>|t|$. 
 
\begin{figure}
\includegraphics[height=240pt,angle=-90]{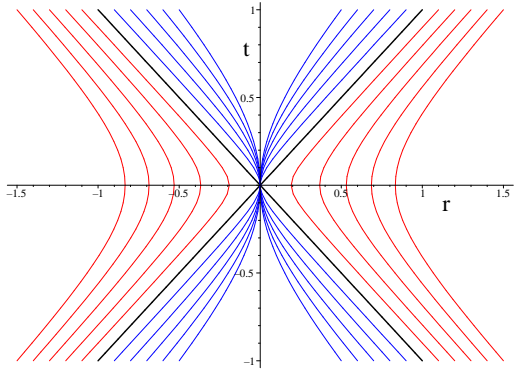}
\caption{\label{fig:0} Fluid trajectories of the $\lambda=2$ exact
solution.}
\end{figure}

The constants $d$, $\lambda$ and $\kappa$ are constrained: In $1+1$ dimensions, $\lambda$ is arbitrary but $\kappa=d=1$, except if $\lambda = 1$ when $\kappa $ is arbitrary.
If $d$ is arbitrary, $\lambda=2$ and $\kappa=d$ is a valid solution both for $|r|<|t|$ and $|r|>|t|$,
and for $\lambda=1$, we have solution inside 
the lightcone for arbitrary $\kappa$.
Recently, T. S. Bir\'o pointed out~\cite{Biro:2007pr}, that 
the $\lambda=1/2$, $\kappa=1$ and the $\lambda=3/2$,
$\kappa=11/3$ cases are solutions for $d=3$ dimensional expansions,
which can be generalized easily for any  $d\in\mathbb{R}$, choosing
$\kappa=(4d-1)/3$.
The function $\phi(\lambda) = 0$ for $\lambda = 0$, 1 or 2,
while it is unity, $\phi(\lambda) = 1$ for $\lambda = 1/2$ or $\lambda = 3/2$. 
It is introduced to indicate, that the solutions are dependent 
on space-time rapidity   $\eta$ in the $\lambda = 1/2$ and $3/2$ cases only.
The $\lambda\neq 1$ solutions describe a flow with relativistic acceleration.
%The $\lambda=1$\,, $|r|>|t|$ case can be interpreted as the
%external Hwa-Bjorken solution, which is also accelerating.
%The trajectory of a fluid element is
%\bl{Bjorkenmotion}
%R(t)=\rec{a_0}\sqrt{1+(a_0 t)^2} \qquad a_0=\rec{r_0}
%\ee
%with $r_0$ initial condition at $t=0$. Each one of these trajectories has a
%constant $a_0$ acceleration in the momentary rest frame.
If $\lambda=2$, the trajectories are uniformly accelerating, 
%with different $a_0$ value, 
see Fig.~\ref{fig:0} and ref.~\cite{Csorgo:2006ax}.

\section{Rapidity distributions and energy density estimation}
\begin{figure}
\includegraphics[width=200pt]{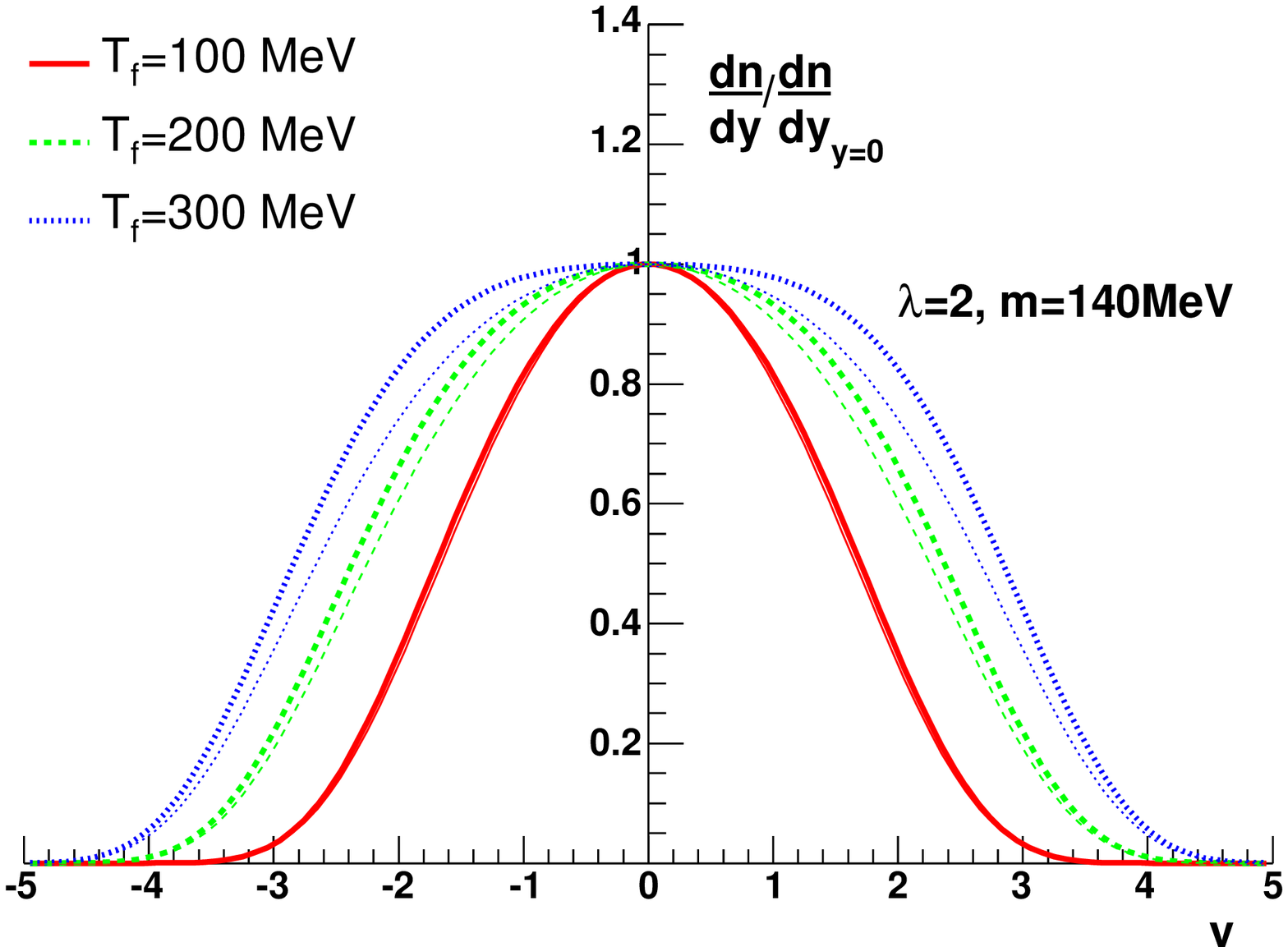}
\includegraphics[width=200pt]{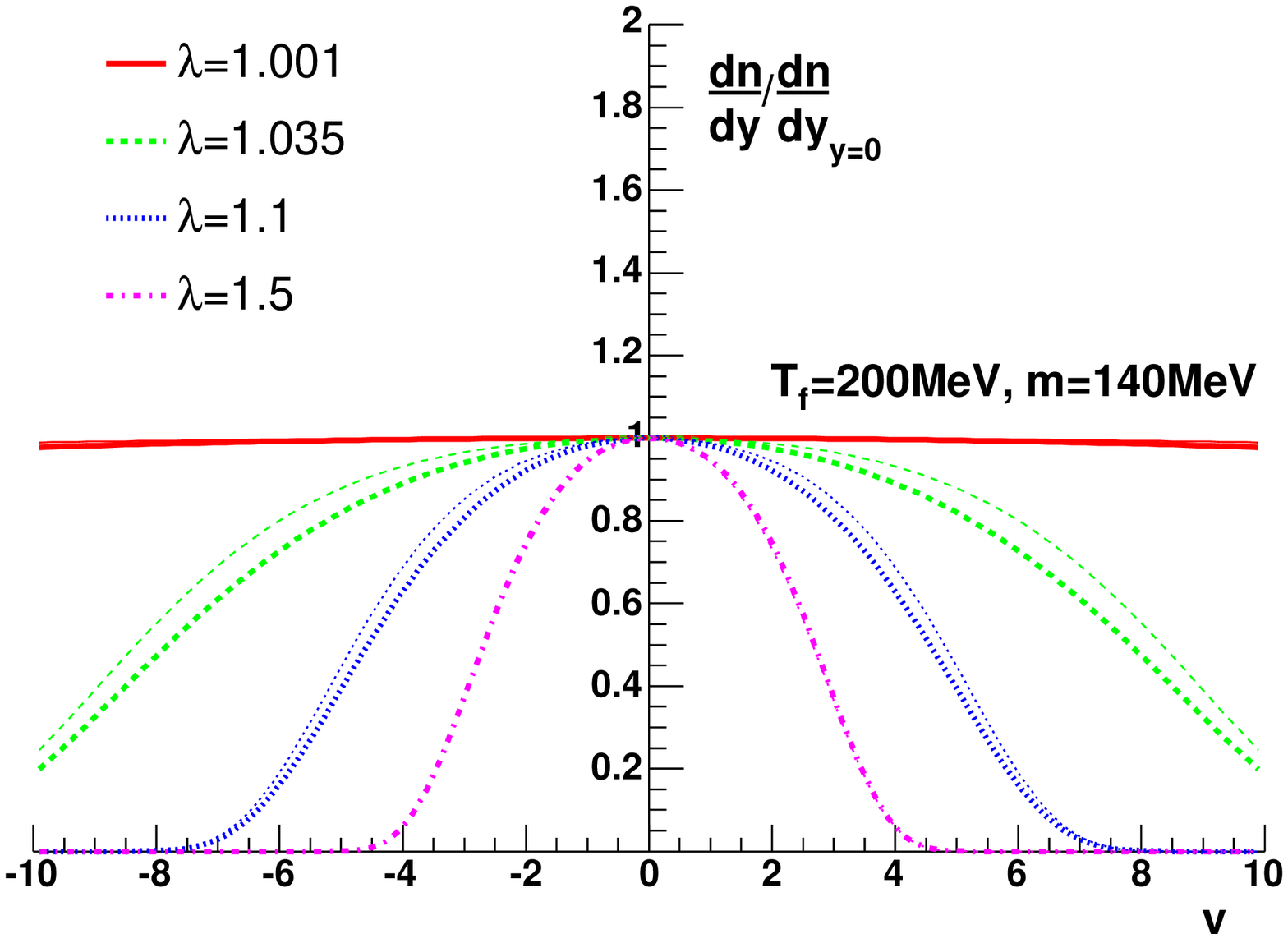}
\includegraphics[width=200pt]{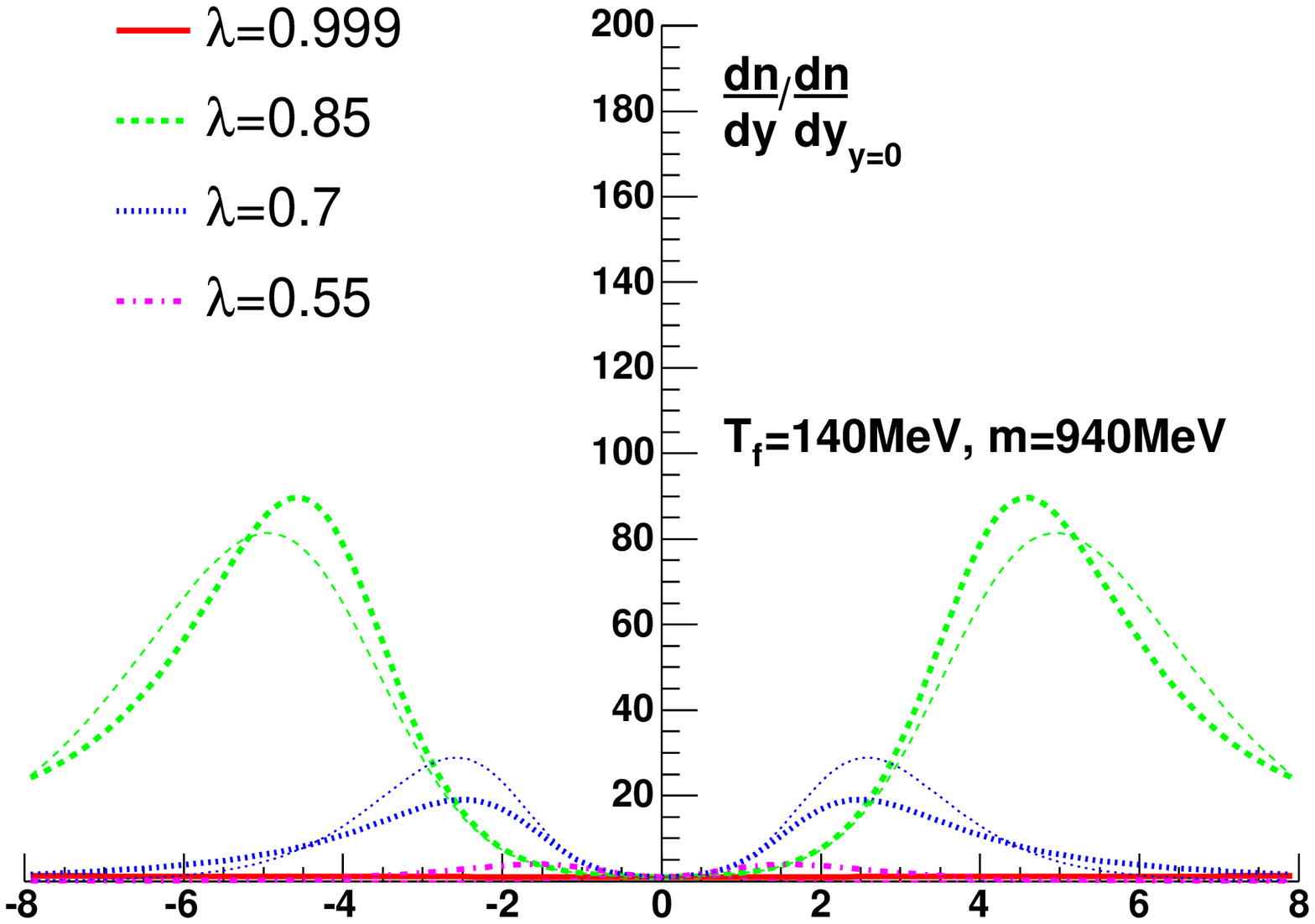}
\caption{\label{fig:1} Normalized rapidity distributions from the new solutions for various
$\lambda$, $T_f$ and $m$ values. Thick lines
show the result of numerical integration, thin lines
the analytic approximation from eq.~\r{e:dndy-approx}. For $\lambda>1$
and not too big $T_f$ it can be used with about 10 \% error. }
\end{figure}

For  $d=1$, the rapidity distribution, $\td{n}{y}$ 
%is given by
%\bl{dist}
%\td{n}{y} =
%    \frac{g}{ 2\pi\hbar }
%    \int p^{\mu}\ud\Sigma_{\mu}(x)
%    \exp\z{\frac{\mu(x)}{T(x)}-\frac{p^{\mu}u_{\mu}(x)}{T(x)}} ,
%\ee
%with $y$ being the rapidity, $\ud\Sigma_{\mu}$ the Cooper-Frye flux term, $p^{\mu}$ the momentum and $g$ the spin-degeneracy factor. Our
was calculated
in a  Boltzmann approximation in ref.~\cite{Csorgo:2006ax}. The freeze-out temperature is  
 $T(\eta=0,\tau=\tau_f ) = T_f$. 
We assumed~\cite{Csorgo:2006ax}, that
the freeze-out hypersurface is pseudo-orthogonal to $u^{\mu}$.
%The equation of this hypersurface is: $\cosh\z{\z{\lambda-1}\eta}=\z{\tau/\tau_f}^{\lambda -1}$.
With a saddle-point integration, for $\lambda > 0.5$, and $m/T_f \gg 1$ (where $m$ is the particle mass),
we got 
\bl{e:dndy-approx}
\td{n}{y}\approx n_f\tau_f
\sqrt{\frac{2\pi m}{T_f\lambda(2\lambda-1)}}\cosh^{\frac{\alpha}{2}-1}\z{\frac{y}{\alpha}}
e^{-\frac{m}{T_f}\cosh^\alpha\z{\frac{y}{\alpha}}} ,
\ee
with $\alpha=\frac{2\lambda-1}{\lambda-1}$.
The ``width'' of this distribution is
\bl{Deta}
    \Delta y^2 = \frac{\alpha }{m/T_f - 1/2 + 1/\alpha} .
\ee
%%%%%%%%%%%%%%%%%%%%%%%%%%%%%%%%%%%%%%%%%%% Fig. 1
The rapidity distribution has a  minimum at $y=0$, if $\Delta y^2<0$ (i.e. $1/2+T_f/(4 m)<\lambda<1$), it
is flat if $\lambda = 1$ or $\lambda = 1/2 + T_f/(4 m)$, otherwise it is approximately Gaussian.
The typical cases are plotted in Fig.~\ref{fig:1}.
%%%%%%%%%%%%%%%%%%%%%%%%%%%%%%%%%%%%%%%%%%%% Fig. 2

\begin{figure}[htb]
\includegraphics[width=200pt]{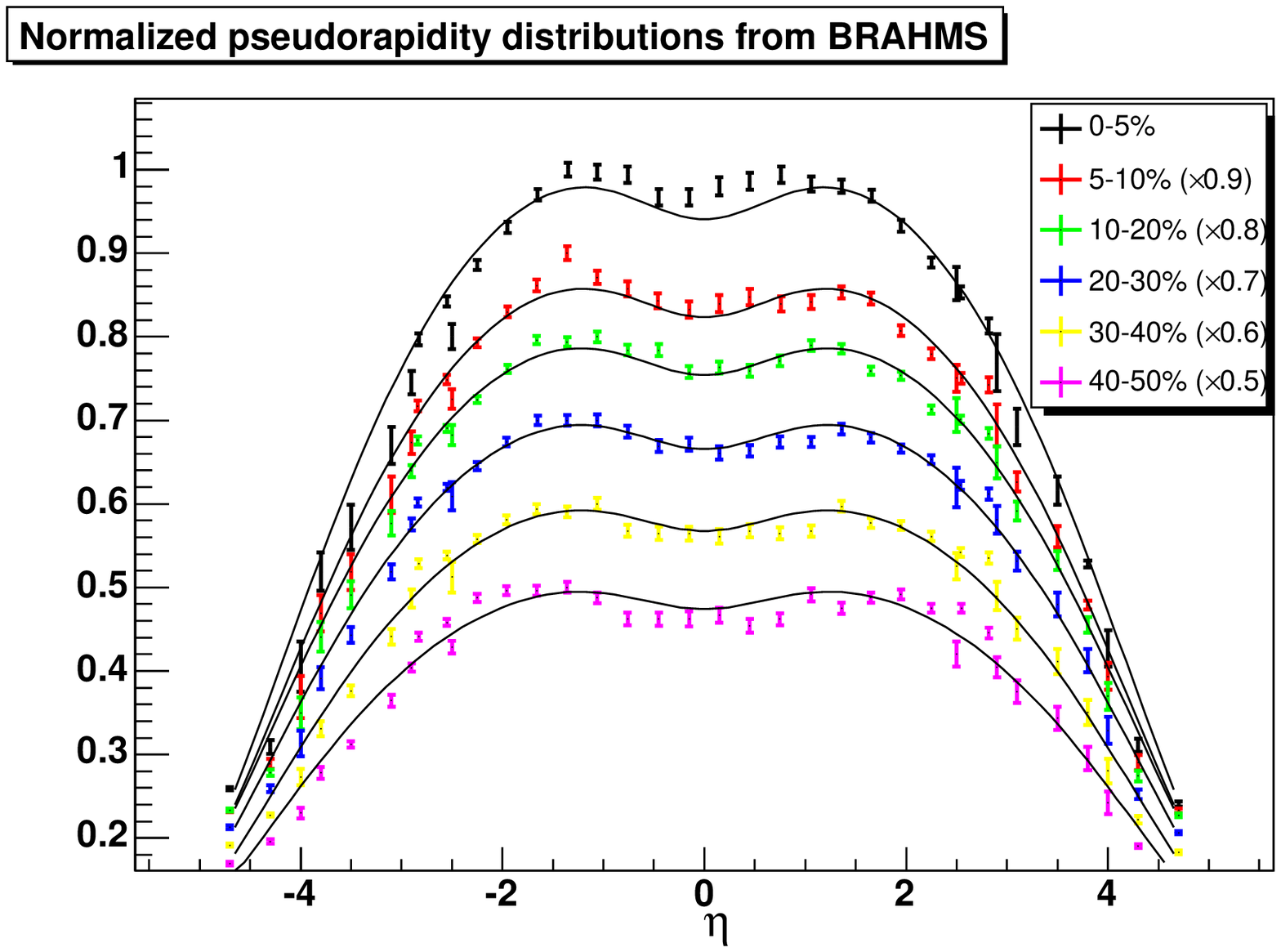}
\includegraphics[width=200pt]{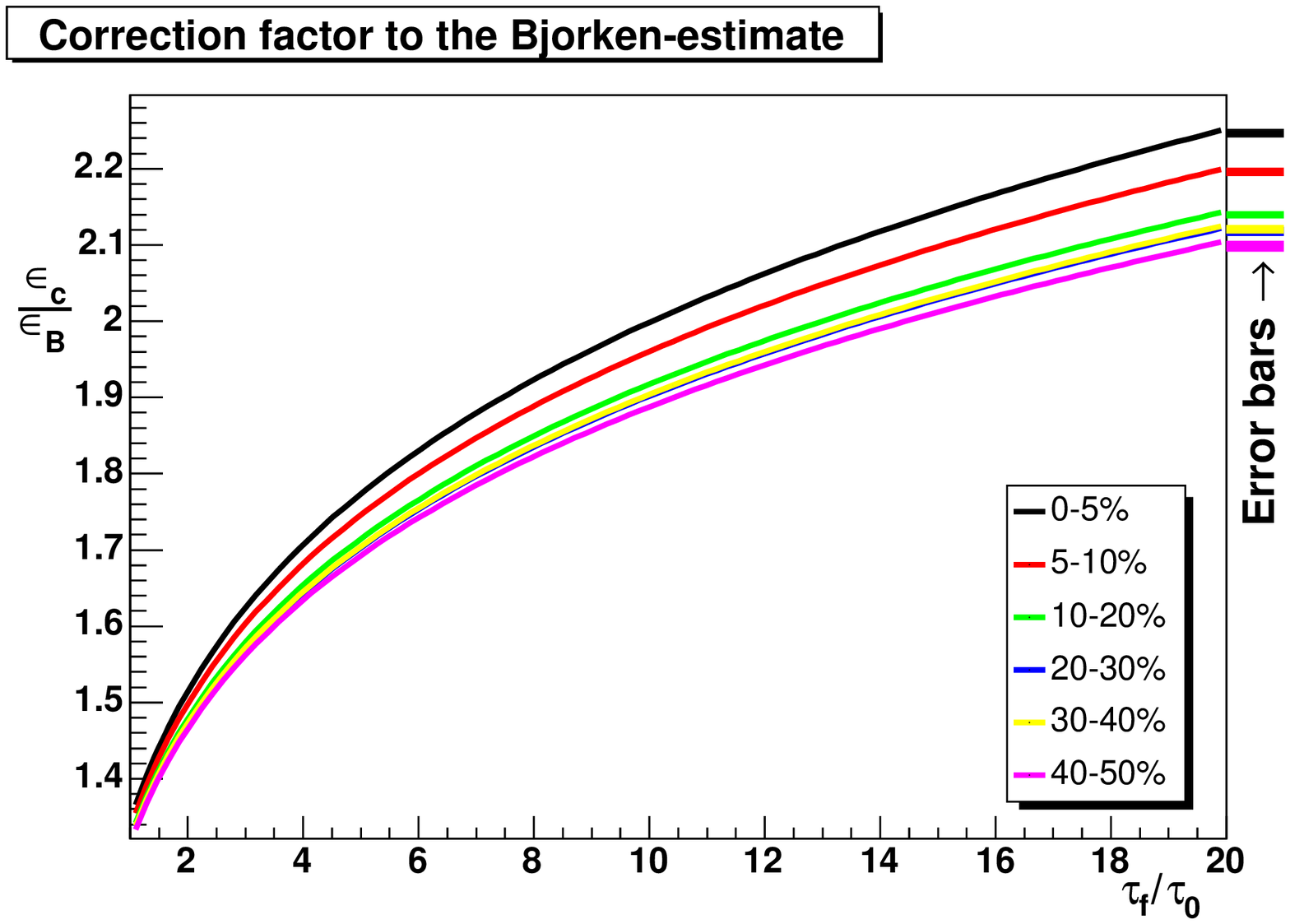}
\caption{\label{fig:3}
Top panel: Fits of eq.~\r{e:dndy-approx} to $dn/d\eta$ data as measured by
the BRAHMS collaboration~\cite{Bearden:2001qq} in $\sqrt{s_{NN}}=200$ GeV Au+Au collisions at various centralities.
Bottom panel: $\varepsilon_c/\varepsilon_B$ ratio as a function of $\tau_f/\tau_0$. Using the Bjorken estimate of
$\varepsilon_B = 5$ GeV/fm$^3$ as given in the BRAHMS White Paper~\cite{BRAHMS-White},
and $\tau_f/\tau_0$=7-10, we find an initial energy density of $\varepsilon_c = (1.7-2.0)\varepsilon_B=8.5-10$ GeV/fm$^3$.}
\end{figure}

As an application, we estimate the energy density reached in heavy ion reactions.
Let us focus on the thin transverse slab at mid-rapidity,
just after thermalization ($\tau=\tau_0$), illustrated by Fig. 2
of ref.~\cite{Bjorken:1982qr}. The radius $R$ of this slab is estimated by the
radius of the colliding hadrons or nuclei, and the volume is $dV=(R^2\pi)\tau_0 d\eta_0$.
The energy content is $dE = \langle m_t\rangle dn$, where $\langle m_t \rangle$ is the average
transverse mass at $y=0$, so similarly to Bjorken, the initial energy density is
\bl{e:Bjorken}
    \varepsilon_0 = \frac{\langle m_t\rangle}{ (R^2 \pi)  \tau_0 }
    \frac{dn}{d\eta_0} .
\ee
For accelerationless, boost-invariant Hwa-Bjorken
flows $\eta_0 = \eta_f = y$, however, for our accelerating solution we have to apply a correction factor of
$\frac{\partial \eta_f}{\partial \eta_0}\,\frac{\partial y}{\partial \eta_f}=\z{\tau_f/\tau_0}^{\lambda -1}\z{2\lambda-1}$.
Thus the initial energy density $\varepsilon_0$ can be
accessed by a corrected  estimation $\varepsilon_c$ as 
\bl{e:ncscs}
\frac{\varepsilon_c}{\varepsilon_B}=\z{2\lambda-1}\z{\frac{\tau_f}{\tau_0}}^{\lambda-1}\,,\,\qquad 
\varepsilon_{B}=\frac{\langle m_t\rangle}{(R^2\pi)\tau_0}\frac{dn}{dy} .
\ee
Here $\varepsilon_{B}$ is the Bjorken estimation, which is recovered if $\td{n}{y}$ is flat (i.e. $\lambda=1$), but for
$\lambda>1$, both correction factors are bigger than 1. Hence the initial energy densities are under-estimated by the Bjorken
formula. Fig.~\ref{fig:3} indicates fits to BRAHMS pseudo-rapidity
distributions from ref.~\cite{Bearden:2001qq}, these fits indicate that $\varepsilon_c=8.5-10$ GeV/fm$^3$ in
Au+Au collisions at RHIC.

\section{Life-time determination}
For a Hwa-Bjorken type of accelerationless, coasting  longitudinal flow,
Sinyukov and Makhlin~\cite{Makhlin:1987gm}
determined the longitudinal length of homogeneity as
\bl{Rlong-b}
    R_{{\rm long}} = \sqrt{\frac{T_f}{m_t}}
        \tau_{B} .
\ee
Here $m_t$ is the transverse mass and
$\tau_{B}$ is the (Bjorken) freeze-out time.
However, acceleration influences the estimated life-times of the reaction.
If the flow is accelerating, the estimated origin of the trajectories is
shifted, and we found the following correction of eq.~\r{Rlong-b},
at mid-rapidity, for broad but finite rapidity distribution:
\bl{Rlong-c}
R_{{\rm long}}=\sqrt{\frac{T_f}{m_t}}\frac{\tau_{{\rm c}}}{\lambda} 
	\quad\Rightarrow\quad 
	\tau_{{\rm c}} = \lambda \tau_{B} .
%\null\vspace{-0.3cm}
\ee
Thus in earlier HBT analyses the effective proper-time duration of the
reaction  have  been \emph{underestimated},
%the smaller is the width of the rapidity distribution, the bigger is the correction factor,
as pointed out also in refs.~\cite{Wiedemann:1999ev,Csanad:2004cj,Renk:2003gn,Pratt:2006jj}.
BRAHMS pseudo-rapidity distributions in Fig.~\ref{fig:3} yield
%measured by the BRAHMS collaboration in $\sqrt{s_{NN}}=200$ GeV Au+Au collisions,
$\lambda \approx 1.2$, 
and imply a 20 \% increase in the estimated life-time of the reaction.

\null\vspace{0.3cm}
\section{Summary}
We have presented new, simple and accelerating solutions of 
relativistic hydrodynamics.
We have improved quantitatively on Bjorken's initial energy density estimate, 
by taking into account the longitudinal work, which is important 
for finite rapidity distributions. 
We have fitted the BRAHMS pseudo-rapidity distributions with the resulting 
simple forms, and pointed
out that in Au+Au collisions at RHIC, 8.5 - 10 GeV/fm$^3$ initial 
energy densities are reached, a factor of 2 larger, than the Bjorken estimate.
We have also corrected the Sinyukov-Makhlin formula
for longitudinal work effects, and found an increase of 
the life-time of the reaction by about 20 \%,
as extracted from the longitudinal HBT radius parameter in Au+Au collisions
at $\sqrt{s_{NN} } = 200 $ GeV.

%\section{Acknowledgements}
{\it Acknowledgements:}
T. Cs. would like to thank Y. Hama and S. S. Padula for
their kind hospitality in Brazil
and for their providing an excellent working atmosphere
during the ISMD 2006 and WPCF2006 conferences.
This work has been supported by the Hungarian OTKA grant T049466,
and by a FAPESP grant from S\~ao Paulo, Brazil.

\end{document}